# A Modified Theoretical Model to Predict the Thermal Interface Conductance Considering Interface Roughness

Yingying Zhang[1#], Dengke Ma[1#], Yi Zang[1], and Nuo Yang[1,2]

[1]State Key Laboratory of Coal Combustion, Huazhong University of Science and Technology (HUST), Wuhan 430074, P. R. China

[2]Nano Interface Center for Energy (NICE), School of Energy and Power Engineering, Huazhong University of Science and Technology (HUST), Wuhan 430074, P. R. China

[#]Y. Z. and D. M. contributed equally to this work.

Corresponding to: nuo@hust.edu.cn (N.Y)

# ABSTRACT

The acoustic mismatch model and the diffuse mismatch model have been widely used to predict the thermal interface conductance. However, the acoustic mismatch model (diffuse mismatch model) is based on the hypothesis of a perfectly smooth (completely disordered) interface. Here, we present a new modified model, named as the mixed mismatch model, which considers the roughness/bonding at the interface. By taking partially specular and partially diffuse transmissions into account, the mixed mismatch model can predict the thermal interface conductance with arbitrary roughness. The proportions of specular and diffuse transmission are determined by the interface roughness which is described by the interfacial density of states. It shows that the predicted results of the mixed mismatch model match well with the values of molecular dynamics simulation and experimental data.

Thermal transport across interface is an important issue for microelectronics, photonics, and thermoelectric devices, and has been studied both experimentally and theoretically recently.[1,2] Generally, the thermal interface conductance (TIC) is used to evaluate the physical properties of thermal transport in devices and materials[3], such as composites[4], superlattices[5], thin-film multilayers[6], nanoscale devices[7], and nanocrystalline materials[8]. Therefore, a deep understanding and an accurate prediction of TIC are crucial to improve the performance of a diverse of devices and materials.

So far, two theoretical models, the acoustic mismatch model (AMM) and the diffuse mismatch model (DMM), have been widely used in predicting the TIC[1]. The AMM makes an essential simplified assumption that phonons incident at an interface undergo specular transmission and are governed by continuum mechanics[1]. Contrarily, the DMM assumes that the interface is completely disordered and phonons lose their memory after reaching the interface.[1,9] Although these two models have advanced the understanding of thermal transport across interface, both the AMM and DMM purely consider the bulk properties, and ignore the interfacial roughness, phonon states[10] and inelastic scattering, which leads to the inaccuracy of the two models.[1,11]

In the past decades, some works have improved the accuracy of the AMM and DMM, and sophisticated modifications to the original models that account for the inelastic scattering have been proposed. For example, the maximum transmission model[12], the higher harmonic inelastic model[13], the joint frequency diffuse mismatch model[14], the scattering-mediated acoustic mismatch model[15] and the anharmonic inelastic model[16]. And the electron-phonon couplings are included when studying the metal/non-metal interface.[17]

Chen proposed the partially diffuse and partially specular interface scattering model to predict the thermal conductivity of superlattice structures[18]. The idea is worth to be applied in calculating the TIC. Besides, it is found that the phonon interface states which are localized close to the interfacial region and different from bulk states have a large effect on TIC.[19-23] However, all of the existing theoretical models do not consider the interfacial states. Therefore, there is a great demand for a theoretical model that accounts for interface states in predicting the TIC.

In this letter, we developed a new model to predict the TIC, named as the mixed mismatch model (MMM). Based on partially specular and partially diffuse transmissions, by taking interface states into account, the MMM can predict the TIC with arbitrary roughness and overcomes the shortage of AMM and DMM. Here, we considered the interface states in obtaining the proportions of diffuse and specular transmission. To validate the accuracy of the MMM, we compared the TIC values of MMM to those by MD simulations and experimental measurements.

At the interface between material A and B, an incident phonon with frequency ω and mode *j* can either scatter back or transmit[1]. According to the Landauer formalism[24], we can predict the TIC, G, as

$$G = \frac{1}{4}\sum_j \int_0^{\omega_{A,j}^v} D_{A,j}(\omega) \frac{\partial n(\omega,T)}{\partial T} \hbar\omega v_{A,j} \alpha_{A\to B,j}(\omega) d\omega \tag{1}$$

where ω is the frequency, $\omega^v$ is the cutoff frequency, D is the phonon density of states, n(ω,T) is the Bose-Einstein distribution function, *v* is the phonon group velocity, and α is the phonon transmission coefficient. The subscript "*j*" refers to the phonon polarization.

In the AMM and DMM, the transmission coefficient can be calculated as[1,6]

$$\alpha_{AMM,A\to B} = \frac{4\rho_A v_A \rho_B v_B}{(\rho_A v_A + \rho_B v_B)^2} \tag{2}$$

$$\alpha_{DMM,A\to B} = \frac{\sum_j D_{B,j} v_{B,j}}{\sum_j D_{A,j} v_{A,j} + \sum_j D_{B,j} v_{B,j}} \tag{3}$$

where *ρ* is the mass density.

According to previous studies[18,25-29], for solid-solid interfaces, both diffuse and specular transmissions are taking place at the interface. Hence, an assumed specular parameter, *p*, can be used to represent the proportion of phonons specular transmitted across the interface[18,25]. So, in the MMM, we define the transmission coefficient by mixing AMM and DMM as

$$\alpha_{DMM,A\to B} = (1-p) \cdot \alpha_{DMM,A\to B} + p \cdot \alpha_{AMM,A\to B} \tag{4}$$

According to Ziman[25], the *p* is related to the root mean square roughness (RMSR), *η*, and phonon wavelength, *λ*, as

$$p = exp\left(-\frac{16\pi^3 \eta^2}{\lambda^2}\right) \tag{5}$$

The interfacial density of states (IDOS) depends on the interfacial structure and RMSR. For the specular interface, the IDOS is as the same as the DOS of crystal structure, where the RMSR is zero. On the other hand, for the diffuse interface, the IDOS is similar to the DOS of an amorphous structure, where the RMSR goes to infinity. We define a relationship between the value of RMSR and the DOS.

$$\eta(\omega) = C \cdot \frac{1}{2} \cdot \left( \frac{D_{int,A}(\omega) - D_{cry,A}(\omega)}{D_{int,A}(\omega) - D_{amo,A}(\omega)} + \frac{D_{int,B}(\omega) - D_{cry,B}(\omega)}{D_{int,B}(\omega) - D_{amo,B}(\omega)} \right) \tag{6}$$

where C is a constant, the $D_{int}$, $D_{amo}$ are the DOS at the interface in crystal and in amorphous structure, respectively.

According to Eq. (5) and Eq. (6), the specular parameter, $p$, is obtained

$$p = exp\left( -\frac{4\pi^3 C^2}{\lambda^2} \cdot \left( \frac{D_{int,A}(\omega) - D_{cry,A}(\omega)}{D_{int,A}(\omega) - D_{amo,A}(\omega)} + \frac{D_{int,B}(\omega) - D_{cry,B}(\omega)}{D_{int,B}(\omega) - D_{amo,B}(\omega)} \right)^2 \right) \tag{7}$$

This is the key equation of this letter. By substituting Eq. (7) into Eq. (4), we can get the transmission coefficient α, then the TIC can be figured out by Eq. (1). It is shown that the interfacial structure/IDOS will greatly affect the phonon transport. As shown in Fig. 3, for the perfectly smooth interface, the specular reflection does not change the energy and the momentum components along the direction of the temperature gradient[18], so the IDOS should be the same as the DOS of crystal structures. Contrarily, for the extremely rough interface, the IDOS is similar to the DOS of amorphous structures. For a partially specular and partially diffuse interface, the IDOS is between the two extreme situations.

For a real interface, the roughness of interfaces could bring two problems: 1) the disorder but continuity of interfacial stress and displacement; 2) the discontinuity of interfacial stress and displacement[30,31]. The specular parameter $p$, however, only considers the continuity part. This will lead to an overvaluation of the MMM prediction to experimental values. So, we introduce a contact coefficient, S, which is less than 1 for a discontinuity interface and equal to 1 for a continuity interface. Thus the measured TIC, G, can be written as

$$G = S \cdot G_{MMM} \tag{8}$$

where $G_{MMM}$ is the value predicted by our model corresponding to a continuity interface.

The MD simulation is an ideal method in predicting the TIC because there is no assumption on the phonon scattering[32]. To demonstrate the accuracy of the MMM, we calculated the TIC of interfaces with different roughness, and then compared the results of MMM with MD values. In our MD simulations, we create two Al/Si interface structures with different roughness, structure (a) and (b) (shown in Fig. S1). The structures after relaxation are shown in Fig. 1(a) and Fig. 1(b). The interface roughness of structure (b) is larger than that of structure (a). And the IDOS of structure (a) and (b) is calculated as the average phonon modes of atoms at the interface. And the interface thickness is set as 0.815nm. We also simulate an amorphous (completely disordered) structure (shown in Fig. 1(c)) to obtain the $D_{amo,Al}$ and $D_{amo,Si}$ in Eq. (8). The MD simulation details are shown in Table S1.

The main results are shown in Fig. 2(a). The TIC of AMM and DMM correspond to the TIC of the perfectly smooth interface and completely disordered interface, respectively. The interfaces of structure (a) and (b) are partially diffuse and partially specular. We calculated the TIC of the two interfacial structures (structure (a) with smaller roughness and structure (b) with larger roughness) by MD. The results of MD show that both of them fall in AMM and DMM predictions, and the TIC of structure (b) is larger than that of structure (a). In the calculation of MMM, the value of C is unknown, but can be obtained by fitting to the MD data of structure (a), as $1.024 \times 10^{-11}$. Then we applied Eq. (7) to predict TIC of structure (b) by MMM which matches well with the data of MD (Fig. 2(a)). It declares the validity of MMM.

The specular parameter is the most important parameter in the MMM which presents the proportion of phonons specular transmitted across the interface. As shown in Fig. 2(b), the dependence of specular parameters on frequency for difference interface structures was calculated and compared. There is an obvious decline in the p-curves of both structure (a) and (b). For lower frequency phonons in each p-curve, with longer wavelength, it shows that the values of p for both (a) and (b) are close to one. That is, the phonons with longer wavelength are not sensitive to the nanoscale interface roughness and transit across them similar to a specular interface. For higher frequency phonons, the shorter wavelength phonons transmit across the interface with more diffusive scatterings. Besides, the decline frequency of (a), ~ 6.5 THz, is higher than that of (b), ~ 3.5 THz, because the roughness of structure (a) is smaller than that of structure (b). As shown in Eq. (5) the IDOS dominates the value of p, which makes the MMM consider the interfacial structure. To show the relationship between the IDOS and the interfacial structure, we calculated the IDOS of different interfacial structures. Figure 3(a) and 3(b) show the IDOS of a perfectly smooth interface in AMM and a completely disordered interface in DMM, corresponding to the DOS of bulk and amorphous structure (Fig. S1 (f)), respectively. While Fig. 3(c) and 3(d) show the IDOS of structure (a) and (b), respectively. By comparing the difference of IDOS between Al and Si, we find that the IDOS of Al and Si become similar with each other as the interface roughness increases because the disorder structure at interface diminishes the bulk crystal properties.

To further validate the MMM, we compare the transmission coefficient calculated by the MMM with the experimental results[33]. In Fig. 4 (a), it shows that the transmission coefficients of the MMM match well with the experimental values. Besides, we can see that the DMM underestimates the transmission coefficient of lower frequency phonons because the phonons with longer wavelength are less likely to be scattered. Contrarily, the AMM overestimates the transmission coefficient of high frequency phonons due to the phonons with shorter wavelength are more likely to be scattered. On the other side, when comparing the value of TIC, the MMM should be higher than the experimental values as explained in Eq. (8) (shown in Fig. 4(b)). To get the parameter S, we fitted the TIC of structure (a) calculated by MMM to the experimental value[33] at 300 K. Then, we use the value of S to obtain the TIC at 350 K and 400 K, respectively, which matches well with the corresponding experimental value.

In summary, we have developed a new theoretical model, named as the mixed mismatch model, which predicts the thermal interface conductance by considering the interfacial structure. The MMM takes partially specular and partially diffuse transmissions into account using the IDOS to determine the proportion of specular parameter. The value of specular parameter has an obvious dependence on the phonon wavelength and interfacial roughness. The MMM is validated by comparing its prediction with both MD results and measurements. Both the value of TIC and the transmission coefficient calculated by the MMM agrees well with the values of MD and the experimental data.

## Acknowledgements

N. Y., Y. Z., D. M., and Y. Z. are supported by the National Natural Science Foundation of China (No.51576076). The authors thank the National Supercomputing Center in Tianjin (NSCC-TJ) and Supercomputing Environment of Chinese Academy of Sciences (SECAS) for providing help in computations.

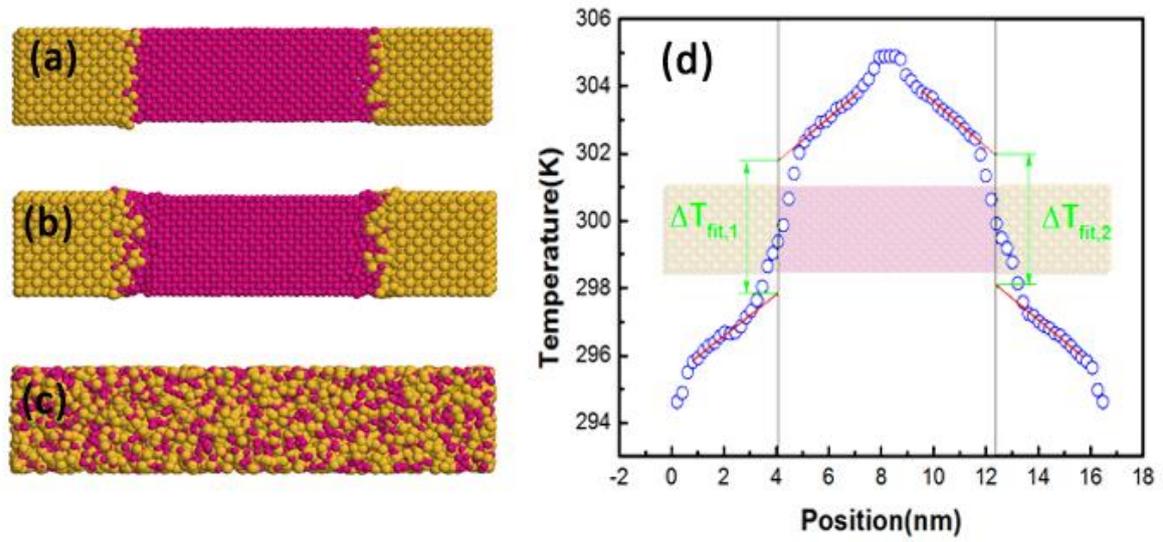

**FIG 1.** (a), (b) and (c) are the sideviews of simulation cell. The red part is Al, and the yellow part is Si. The lattice constants of Al and Si are 0.407 and 0.543nm, respectively. (a) Al 8×8×20unit cells$^3$ /Si 6×6×15unit cells$^3$ is the small interface roughness structure. (b) Al 8×8×20unit cells$^3$ /Si 6×6×15unit cells$^3$ is the large interface roughness structure. (c) This completely disordered structure is used to obtain the IDOS of DMM. (d) A typical time-averaged temperature profile of interface structure. The interface temperature difference is defined as average of the difference between the linearly extrapolated temperature at each side of the interface. Then, $\Delta T_{fit}=(\Delta T_{fit,1}+\Delta T_{fit,2})/2$.

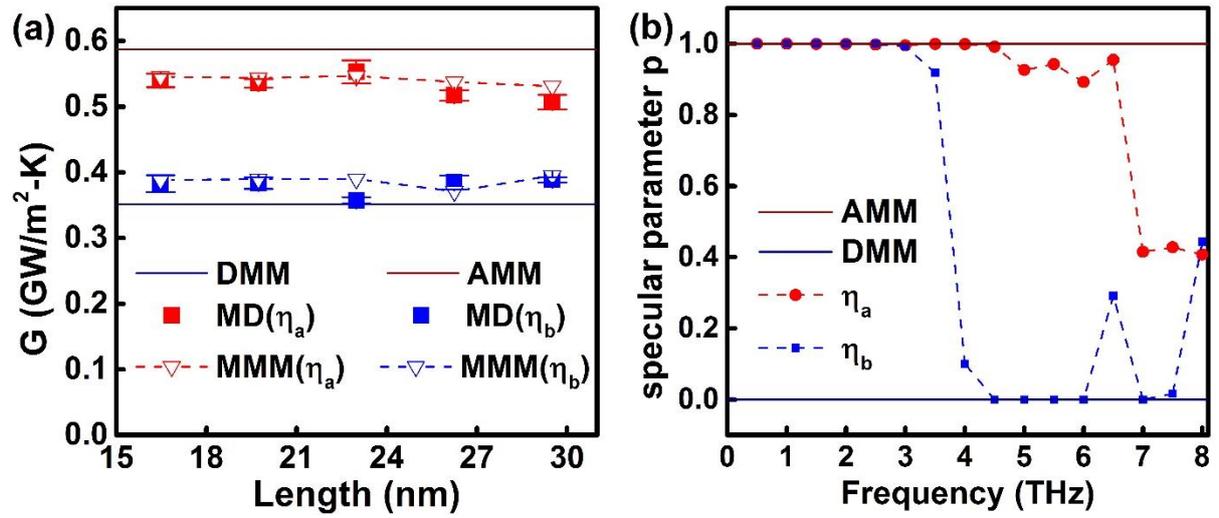

**FIG 2.** (a) The results of thermal interface conductance calculated by AMM, DMM, and MMM. MD means our MD calculation results of TIC. $\eta_a$ and $\eta_b$ denote different interface roughness, and the roughness of $\eta_a$ is smaller than that of $\eta_b$; (b) The specular parameter $p$ of structures with different interface roughness: $\eta_a$ and $\eta_b$, which correspond with the structures in Fig. 1(a) and Fig. 1(b).

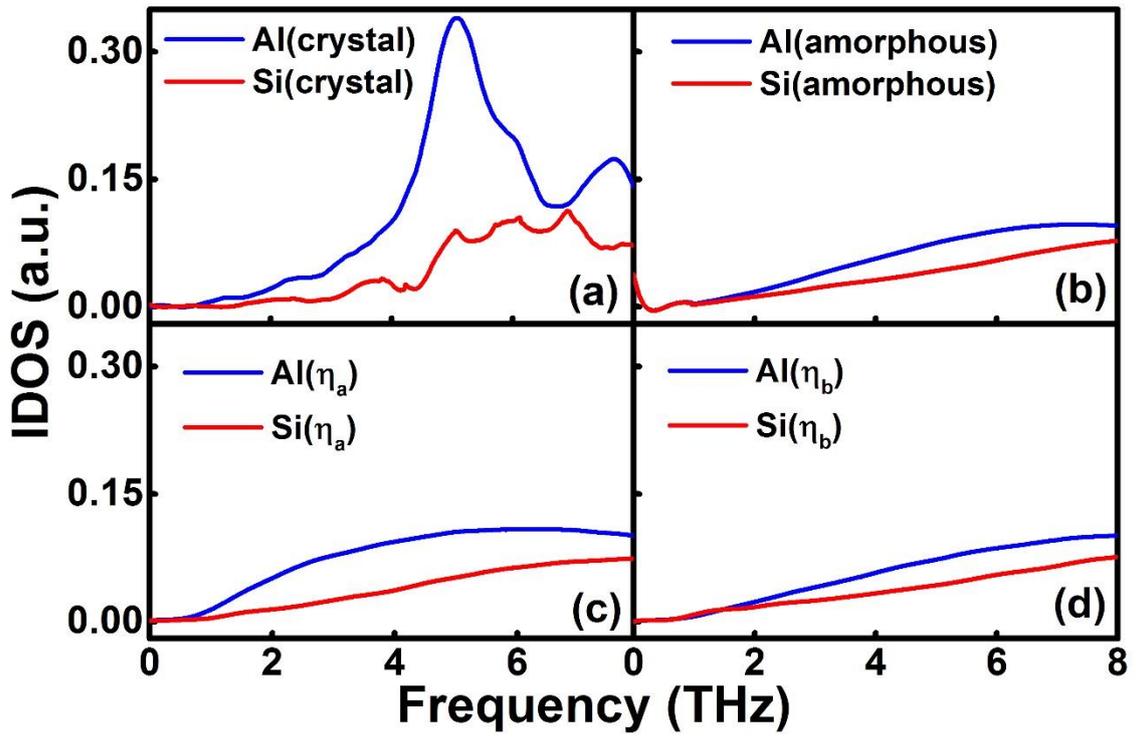

**FIG 3.** (a) DOS of Al and Si in crystal structure; (b) DOS of Al and Si in amorphous structure; (c) IDOS of structure (a) with a system length as 16.473 nm; (d) IDOS of structure (b) with a system length as 16.473 nm.

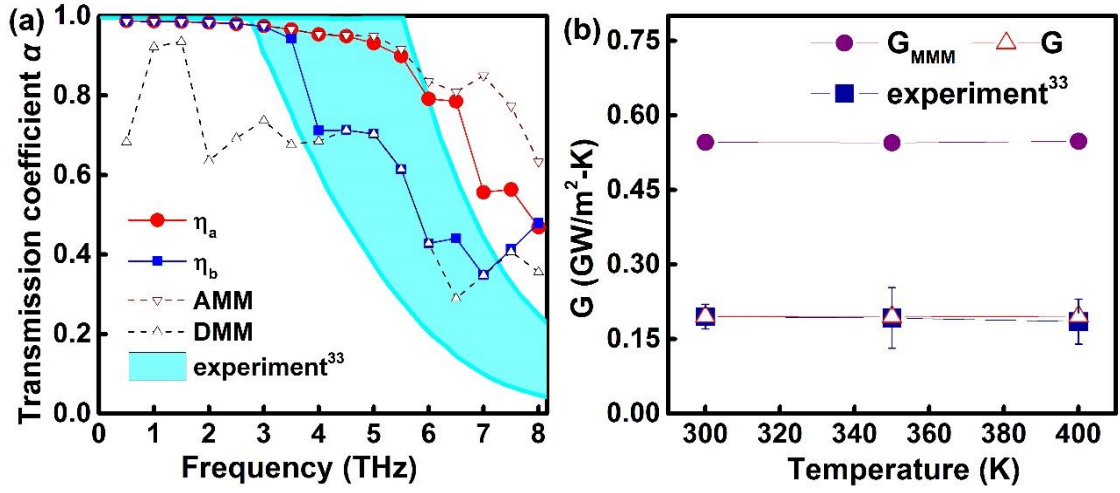

**FIG 4.** (a) Comparing the transmission coefficient α of MMM to that of AMM, DMM and experimental data[33]. (b) $G_{MMM}$ is the TIC predicted by our MMM model. G is obtained from the Eq. (8). And the parameter S is calculated by substituting the value of $G_{MMM}$ and experimental data at 300K for the $G_{MMM}$ and G in Eq. (8). Then we use the value of S to get the G at 350K and 400K, and make a comparison between the G and the experimental measurements.

Supplementary Information

# A Modified Theoretical Model to Predict the Thermal Interface Conductance Considering Interface Roughness

Yingying Zhang[1#], Dengke Ma[1#], Yi Zang[1], and Nuo Yang[1,2]

[1]Nano Interface Center for Energy (NICE), School of Energy and Power Engineering, Huazhong University of Science and Technology (HUST), Wuhan 430074, P. R. China

[2]State Key Laboratory of Coal Combustion, Huazhong University of Science and Technology (HUST), Wuhan 430074, P. R. China

[#]Y. Z. and D. M. contributed equally to this work.

Electronic mail: nuo@hust.edu.cn

# SI. Molecular dynamics simulation details

In our MD simulation, we create different initial interface structures in Al/Si superlattices to get the interfaces with different roughness. Figure. S1 shows the corresponding interface structures of different models. And in table 1, we list the specific form and parameters of our potential function: 2NN MEAM, and show the details in our FORTRAN code.

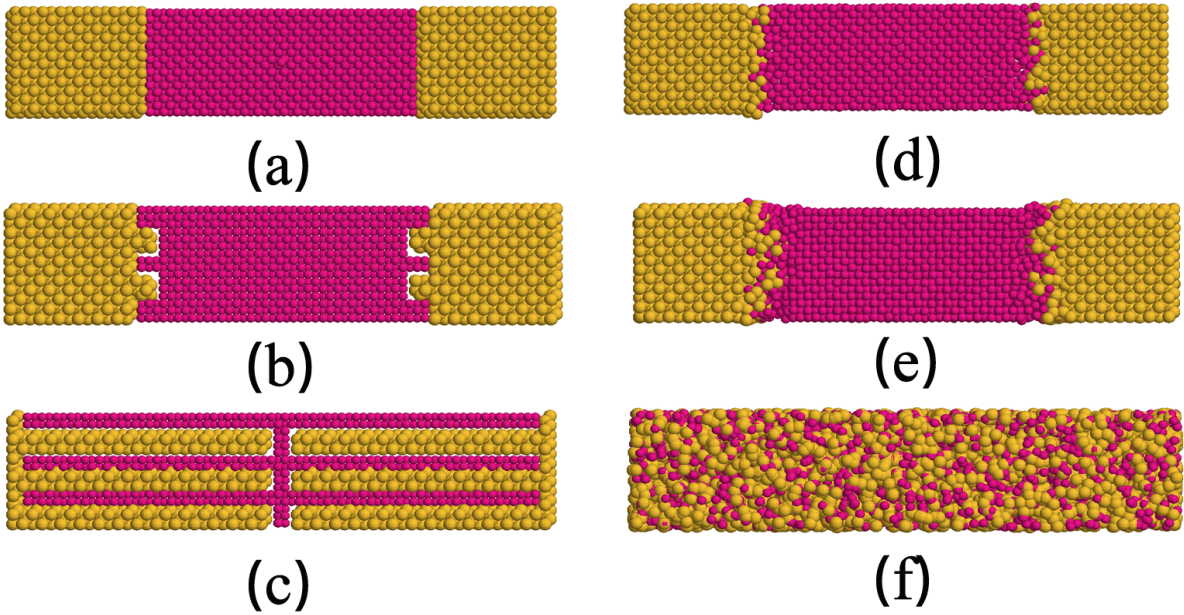

**FIG S1**. The sideviews of simulation cell. The red part is Al, and the yellow part is Si. The lattice constants of Al and Si are 0.407 and 0.543nm, respectively. (a), (b) and (c) are the initial structures of MD simulation, while (d), (e) and (f) are the corresponding stable structures after relaxing process. (d) Al 8×8×20unit cells$^3$ /Si 6×6×15unit cells$^3$ is the small interface roughness structure. (e) Al 8×8×20unit cells$^3$ /Si 6×6×15unit cells$^3$ is the large interface roughness structure. (f) This completely disordered structure is used to obtain the IDOS of DMM.

**Table S1.** MD simulation details and parameters.

| Method | Non-Equilibrium MD (Direct method) | | | | | |
|---|---|---|---|---|---|---|
| **Potential (2NN MEAM)** | | | | | | |
| **Function** | $E = \sum_i \left[ F_i(\overline{\rho}_i) + \frac{1}{2} \sum_{j(\neq i)} \phi_{ij}(R_{ij}) \right]$ | | | | | |
| | $F(\overline{\rho}) = AE_c(\overline{\rho}/\overline{\rho}^0)\ln(\overline{\rho}/\overline{\rho}^0)$ | | | | | |
| | $(\rho_i^{(0)})^2 = \left[ \sum_{j \neq i} \rho_j^{a(0)}(R_{ij}) \right]^2$ | | | | | |
| | $(\rho_i^{(1)})^2 = \sum_\alpha \left[ \sum_{j \neq i} (R_{ij}^\alpha/R_{ij}) \rho_j^{a(1)}(R_{ij}) \right]^2$ | | | | | |
| | $(\rho_i^{(2)})^2 = \sum_{\alpha,\beta} \left[ \sum_{j \neq i} \frac{R_{ij}^\alpha R_{ij}^\beta}{R_{ij}^2} \rho_j^{a(2)}(R_{ij}) \right]^2 - \frac{1}{3} \left[ \sum_{j \neq i} \rho_j^{a(2)}(R_{ij}) \right]^2$ | | | | | |
| | $(\rho_i^{(3)})^2 = \sum_{\alpha,\beta,\gamma} \left[ \sum_{j \neq i} \left( \frac{R_{ij}^\alpha R_{ij}^\beta R_{ij}^\gamma}{R_{ij}^3} \right) \rho_j^{a(3)}(R_{ij}) \right]^2 - \frac{3}{5} \sum_\alpha \left[ \sum_{j \neq i} \left( \frac{R_{ij}^\alpha}{R_{ij}} \right) \rho_j^{a(3)}(R_{ij}) \right]^2$ | | | | | |
| | $\overline{\rho}_i = 2\rho_i^{(0)}/(1+e^{-\Gamma})$ | | | | | |
| | $\Gamma = \sum_{h=1}^3 t_i^{(h)} \left[ \rho_i^{(h)}/\rho_i^{(0)} \right]^2$ | | | | | |
| | $\rho_j^{a(h)}(R) = e^{-\beta^{(h)}(\frac{R}{r_e}-1)}$ | | | | | |
| | $S_{ijk} = f_c[(C-C_{min})/(C_{max}-C_{min})]$ | | | | | |
| **Parameters Al** | $E_C$ | $r_e$ | $B$ | $A$ | $\beta^{(0)}$ | $\beta^{(1)}$ |
| | 3.36 | 2.8613 | 0.4955 | 1.16 | 3.2 | 2.6 |
| | $\beta^{(2)}$ | $\beta^{(3)}$ | $t^{(1)}$ | $t^{(2)}$ | $t^{(3)}$ | $\rho^0$ |
| | 6 | 2.6 | 3.05 | 0.51 | 7.75 | 1 |
| **Parameters** | $E_C$ | $r_e$ | $B$ | $A$ | $\beta^{(0)}$ | $\beta^{(1)}$ |
| | 4.63 | 2.3517 | 0.6191 | 0.58 | 3.55 | 2.5 |

| Si | $\beta^{(2)}$ | $\beta^{(3)}$ | $t^{(1)}$ | $t^{(2)}$ | $t^{(3)}$ | $\rho^0$ |
|---|---|---|---|---|---|---|
| | 0 | 7.5 | 1.8 | 5.25 | -2.61 | 1.88 |
| **Parameters** | Al-Al-Al | | Si-Si-Si | | Al-Si-Al/Si-Al-Si | |
| | $C_{max}$ | $C_{min}$ | $C_{max}$ | $C_{min}$ | $C_{max}$ | $C_{min}$ |
| | 2.8 | 0.49 | 2.8 | 1.41 | 1.44 | 0.14 |

### Simulation process

| Ensemble | Setting | | | | Purpose |
|---|---|---|---|---|---|
| NVT | Runtime (ns) | 0.1 | Unit cell (nm) | | Relax structure |
| | Temperature (K) | 300 | Al | Si | |
| | | | 0.4047 | 0.5431 | |
| | Boundary condition | X, Y, Z: periodic, periodic, periodic | | | |
| NVE | Runtime (ns) | 2.62 | | | Record information |
| | Boundary condition | X, Y, Z: periodic, periodic, periodic | | | |
| | Thermostat | Heat source | 305 K | Al | |
| | | Heat sink | 295 K | Si | |

### Recorded physical quantity

| Temperature | $<E> = \sum_{i=1}^{N} \frac{1}{2} m v_i^2 = \frac{1}{2} N k_B T_{MD}$ |
|---|---|
| **Heat flux** | $J = \frac{1}{N_t} \sum_{i=1}^{N_t} \frac{\Delta \varepsilon_i}{2 \Delta t}$ |
| **Thermal conductance** | $\kappa = -\frac{J}{A \cdot \Delta T}$ |

## SII. The phonon dispersion relationship of Al and Si.

The phonon dispersion relationship was obtained by using the general utility lattice program (GULP). It is shown that the highest frequency of phonon in Si (~ 16 THz) is much larger than those in Al (~ 8 THz). However, the high frequency phonons in Si (> 8 THz) will not contribute to the TIC because the model only takes elastic scattering into account. So in our model, we only consider the overlap frequency of two materials.

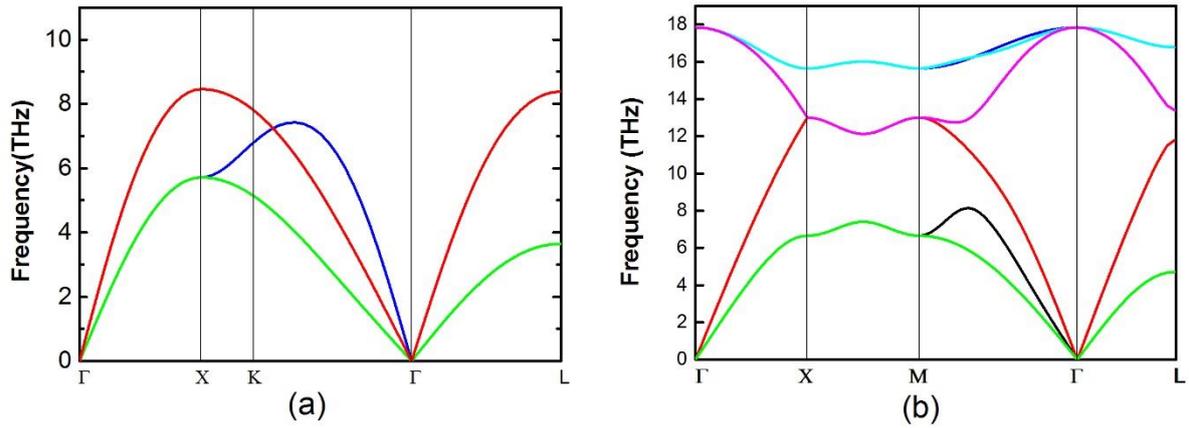

**FIG S2**. Phonon dispersion relationship of (a) Al and (b) Si, respectively. .

**SIII. The interfacial density of states (IDOS) of structures with different system length before frequency 8THz.**

We listed the IDOS of different system length in Fig. S3. In this figure, we can conclude that the DOS of Al and Si are much closer due to the increasing interface roughness.

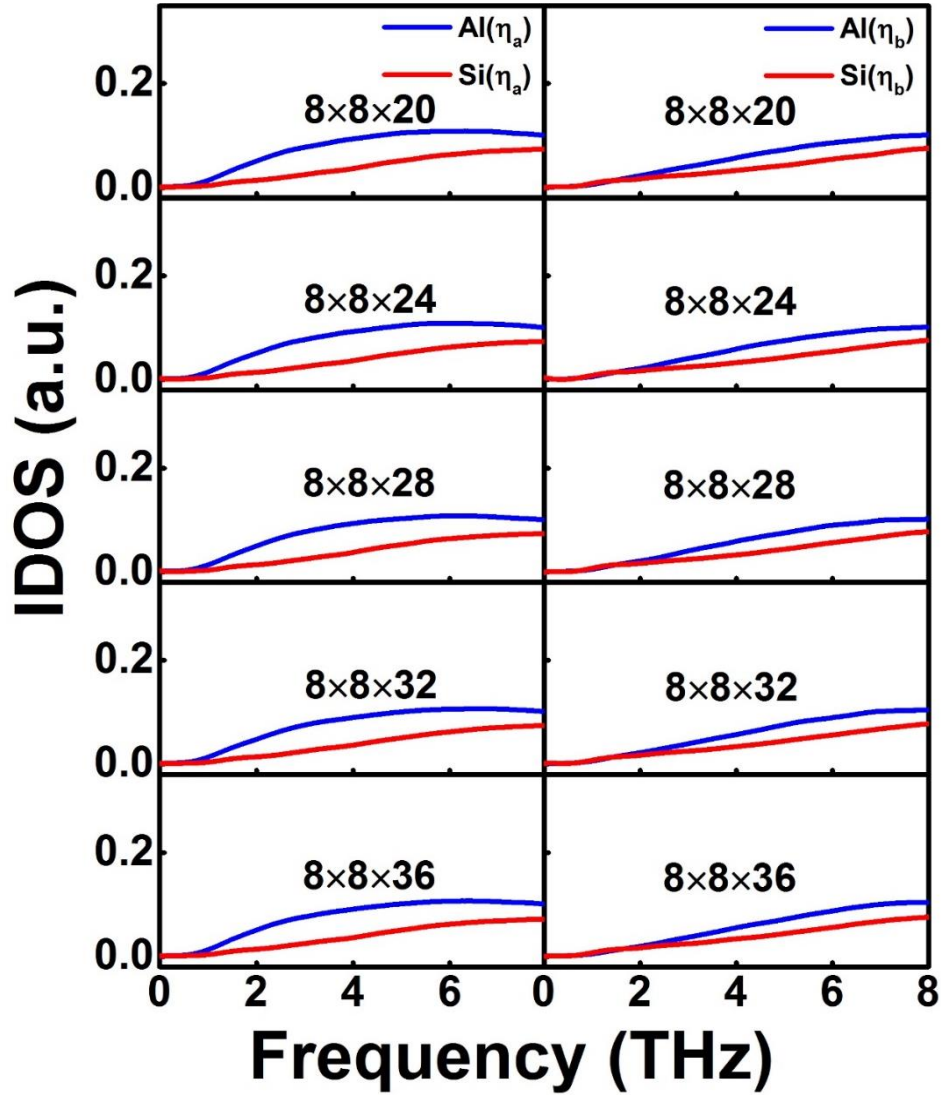

**FIG S3**. The IDOS of structures with different system length. The figures in left column are the IDOS of small interface roughness structures, the figures in right column are the IDOS of large interface roughness structures. 8×8×20 denotes the number of unit cells in xyz directions. And the cut-off frequency in the figures are 8THz.

**SIV. The interfacial density of states (IDOS) of structures with different system length of the overall frequency.**

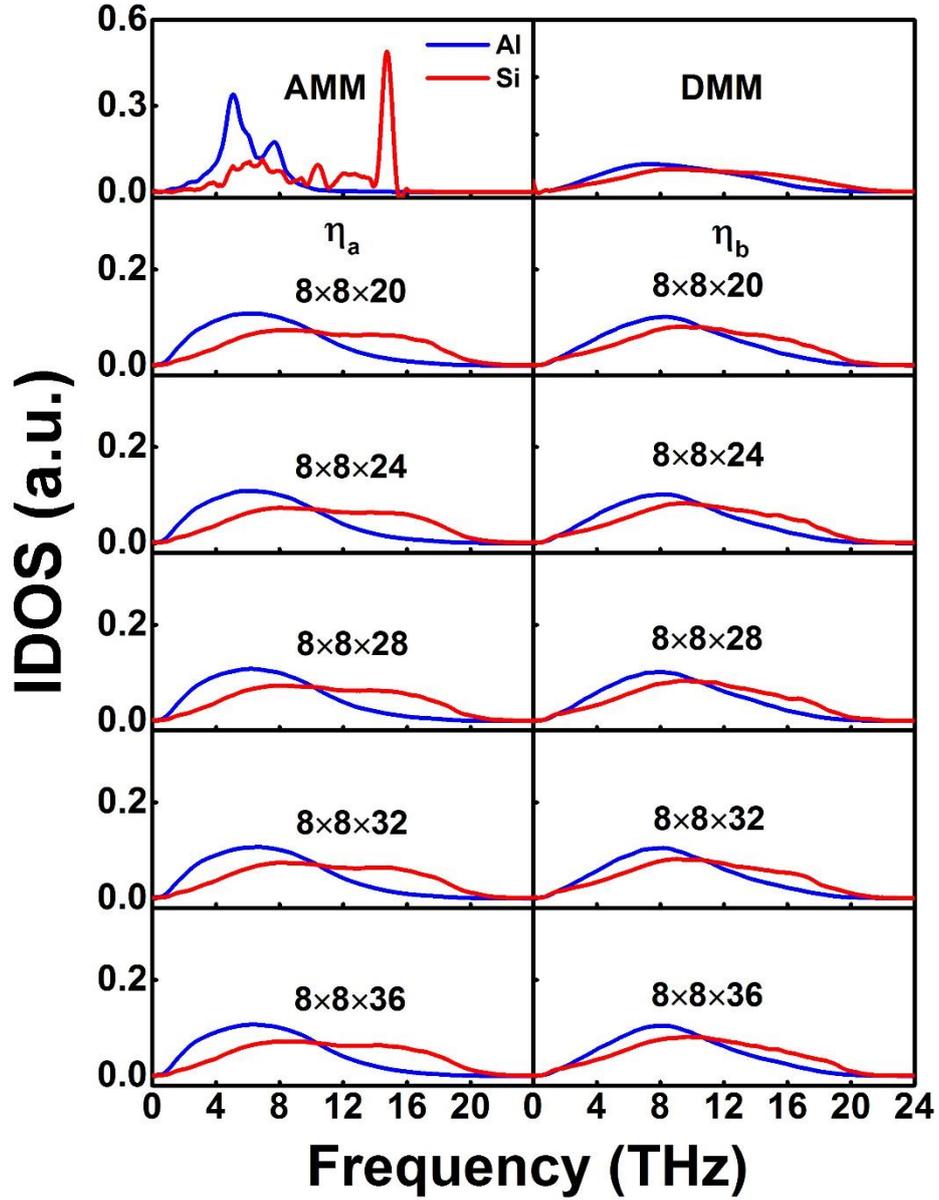

**FIG S4.** The IDOS of the acoustic mismatch model (AMM), diffuse mismatch model (DMM), and structures with different system length. The figures in left column are the IDOS of small interface roughness structures, the figures in right column are the IDOS of large interface roughness structures. 8×8×20 denotes the number of unit cells in xyz directions. All the figures show the IDOS of overall frequency.

## SV. The specular parameter *p* of different interface conditions of different system length.

Figure. S5 shows the specular parameter of different system length under different interface roughness. We could see that the decline frequency of structure b is smaller than that of structure a. When the system unit is $8\times8\times36$, the p of structure b shows an abrupt decline at the frequency 3THz, we still do not have a clear analysis about it recently.

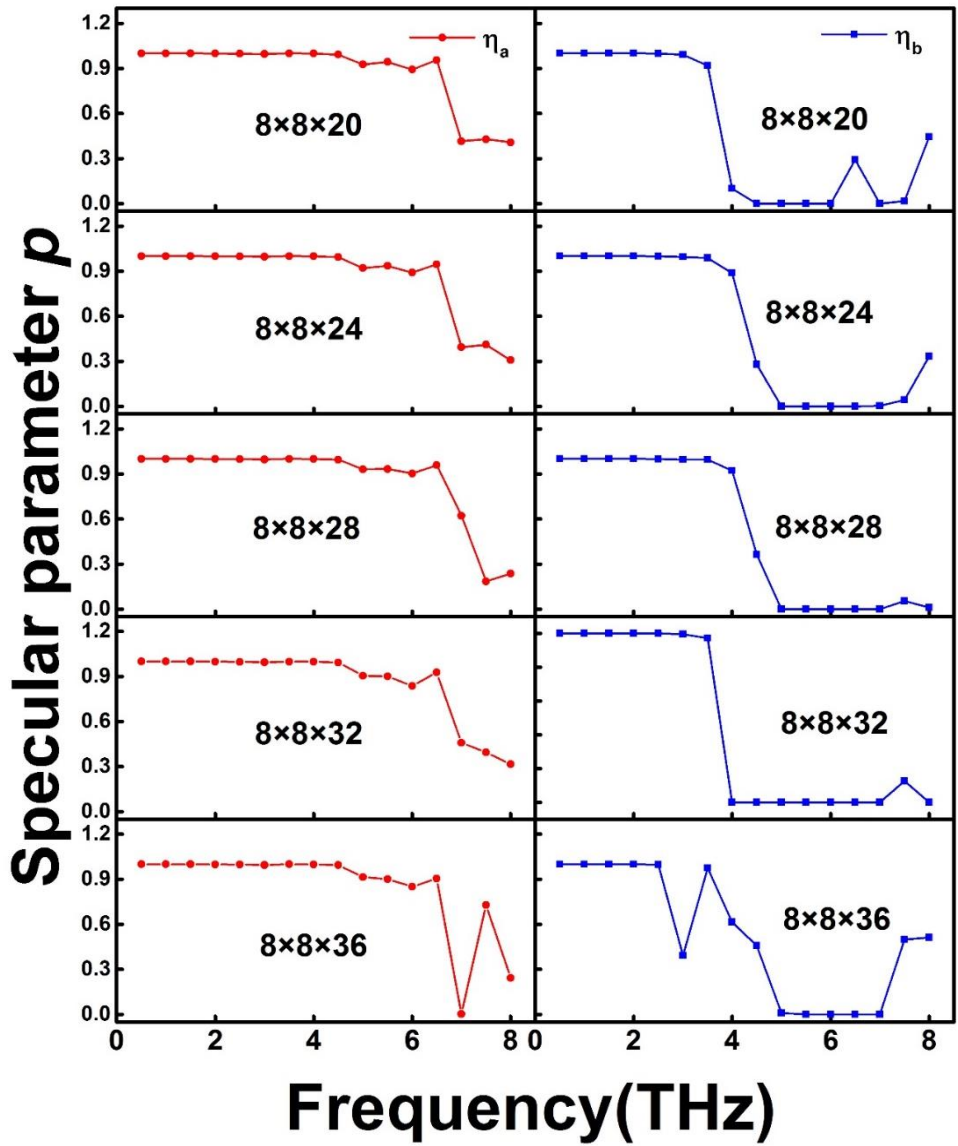

**FIG S5**. The specular parameter *p* of different interface conditions: small roughness ($\eta_a$) and large roughness ($\eta_b$) of interface.